\documentstyle[11pt,aaspptwo]{article} 

\def\fnsiz{\footnotesize}
\def\ts{\thinspace}
\newlength{\zwidth}
\newcommand{\as}{\mbox{$''$}}
\newcommand{\dg}{\mbox{$^{\circ}$}}

\newcommand{\etal}{{\it et~al.\/}}
\newcommand{\gapeq}{\mbox{$~\stackrel{\scriptstyle >}{\scriptstyle \sim}~$}}
\newcommand{\Hline}[1]{\mbox{H{\fnsiz {#1}}}}
\newcommand{\Halpha}{\Hline{\mbox{$\alpha$}}}
\newcommand{\HI}{\mbox {H\thinspace{\fnsiz I}}}
\newcommand{\HII}{\mbox {H\thinspace{\fnsiz II}}}
\newcommand{\ie}{{\it i.e.}}
\newcommand{\kms}{\mbox{km\ts s$^{-1}$}}
\newcommand{\lapeq}{\mbox{$~\stackrel{\scriptstyle <}{\scriptstyle \sim}~$}}
\newcommand{\MHI}{\mbox{${\cal M}_{\rm HI}$}}
\newcommand{\Msun}{\mbox{${\cal M}_\odot$}}
\newcommand{\pino}{\parindent=0mm}

\newcommand{\RHo}{\mbox{$R_{\rm Ho}$}}
\newcommand{\skiggy}{\rm SGC\-0938.1-7623}
\newcommand{\Vrot}{\mbox{$V_{\rm rot}$}}
\newcommand{\Vsys}{\mbox{$V_{sys}$}}
\newcommand{\Wfo}{\mbox{$W_{50}$}}
\newcommand{\Wto}{\mbox{$W_{20}$}}
\newcommand{\zsp}{\mbox{\hspace{\zwidth}}}

\begin{document}
\parskip=1ex

\title{NGC~2915. II. A Dark Spiral Galaxy \\
With A Blue Compact Dwarf Core}

\author{Gerhardt R. Meurer}
\affil{Department of Physics and Astronomy, \\
The Johns Hopkins University \\ 
Baltimore, MD 21218-2695 \\
Electronic-mail: meurer@poutine.pha.jhu.edu}

\author{Claude Carignan}
\affil{D\'epartement de Physique and Observatoire du Mont M\'egantic, \\
Universit\'e de Montr\'eal \\
C.P.\ 6128, Succ.\ ``A'', \\
Montr\'eal, Qu\'ebec, Canada H3C 3J7 \\
Electronic-mail: claude@astro.umontreal.ca}

\author{Sylvie F. Beaulieu and Kenneth C. Freeman}
\affil{Mount Stromlo and Siding Spring Observatories \\
The Australian National University \\
Private Bag, P.O.\ Weston Creek, ACT 2611, Australia \\
Electronic-mail: beaulieu@merlin.anu.edu.au, kcf@merlin.anu.edu.au }

\vskip 0.5cm  

\centerline{Submitted to {\em The Astronomical Journal\/}}

\begin{abstract} This paper presents {\it Australia Telescope Compact
Array\/} \HI\ synthesis observations of the weak blue compact dwarf
(BCD) galaxy NGC~2915.  It is shown that NGC~2915 has the \HI\
properties of a late type spiral galaxy (Sd - Sm), including a double
horn global profile, and \HI\ spiral arms.  The \HI\ extends out to
over five times the Holmberg radius, and 22 times the exponential
scale length in the {\em B\/} band.  The optical counterpart
corresponds to a central \HI\ bar. The \HI\ distribution and
kinematics are discussed in detail.  A rotation curve is derived and
fitted with a mass model consisting of a stellar disk, a neutral gas
disk, and a dark matter (DM) halo.  The DM halo dominates at nearly
all radii.  The total mass to blue light ratio, ${\cal M}_T/L_B = 76$
within the last measured point.  Thus NGC~2915 is one of the darkest disk
galaxies known. The complex \HI\ dynamics of the central region
results in a high uncertainty of many of the fitted parameters.
Nevertheless it is clear that the core of the DM halo is unusually
dense ($\rho_0 \approx 0.1\, \Msun\, {\rm pc^{-3}}$) and compact ($R_c
\approx 1$ kpc).  The neutral gas component, with mass $M_g =
1.27\times 10^9\,\Msun$ is probably more massive than the stellar
disk.  Split and broad \HI\ lines (velocity dispersion $\approx 35\,
\kms$) are seen in the central region.  Pressure support is probably
significant, and it is not clear whether the core is in equilibrium.
Beyond the optical disk the average \HI\ line of sight velocity
dispersion is 8 \kms, which is normal for disk galaxies.  NGC~2915
does not obey the Tully-Fisher (1977) relation, being underluminous
for its $V_{\rm rot} = 88$ \kms\ by a factor of nine.  It also does
not obey the star formation threshold model of Kennicutt (1989), when
only the neutral gas is considered. A simple \HI\ surface density
threshold of $\Sigma_{\rm HI,crit} \approx 10^{21}\, {\rm cm}^{-2}$
adequately describes the location of current star formation.  Although
the \HI\ properties of NGC2915 are extreme relative to normal galaxies
they appear less extreme in comparison to other BCDS, which have
similar radial profiles of \HI\ density and velocity dispersion, and
\HI\ extending well beyond the optical disk.
\end{abstract}

\section{Introduction}\label{s:intro}

The quest for understanding dark matter (DM) halos of galaxies using
\HI\ as a probe is predominantly weighted towards spiral and dwarf
irregular galaxies.  The \HI\ rich blue compact dwarf (BCD;
\markcite{Thuan and Martin, 1981}) class has been largely ignored, and
is under represented in \HI\ synthesis studies.  This may be because
their small optical angular sizes, often smaller than typical
synthesized beam sizes, suggests that they will not be well resolved
at $\lambda$21 cm.  Also, only 8\%\ of the BCDs in Thuan and Martin's
sample have double horn profiles indicative of extended rotating
disks, suggesting that they are not ideal candidates for DM studies.
Some important \HI\ synthesis observations of BCDs, and similar
galaxies, have been made (e.g.\ Taylor \etal, 1993; 1995; 1996; Hunter
\etal, 1994) but their primary aims were other than to search for DM.

Here we present \HI\ observations of NGC~2915 whose optical properties
were presented by \markcite{Meurer \etal\ (1994;  paper I)}. There it
was shown to be a weak BCD, that is its integrated star formation rate
is only 0.05 \Msun\, yr$^{-1}$.  Most of this star formation is near the
very center of the galaxy with some enhanced star formation along the
major axis to the SE of the center.  NGC~2915 resolves into stars; the
brightest of these yields the distance $D = 5.3 \pm 1.3$ Mpc. At this
distance 1\as\ corresponds to 26 pc, and 1$'$ corresponds to 1.54 kpc.
Previous single dish \HI\ observations indicate that it has a very
extended \HI\ disk (Becker \etal\ 1988).  The \HI\ observations
presented here were made primarily to determine NGC~2915's \HI\ rotation
curve, and from it, constrain the structure of its DM halo.  In addition
we wish to examine how the \HI\ properties of this BCD differ from other
types of galaxies, and whether they give any indication of what
regulates star formation in BCDs.  

The observations are presented in \S\ref{s:obs}.  Section
\S\ref{s:morph} presents an overview of the \HI\ properties.  The
dynamics of the ISM are analyzed in \S\ref{s:dyn}.  The rotation curve
is measured, pressure support corrections applied and mass models
fitted.  Section \S\ref{s:disc} presents a discussion of the results
including a comparison with other galaxies, a discussion of the
Tully-Fisher (1977) relation and the star-formation threshold law. Our
conclusions are summarized in \S\ref{s:conc}. In the Appendix we
discuss a possible interaction partner of NGC~2915.

\section{Observations and Data Reduction}\label{s:obs}

NGC~2915 was observed with three configurations of the Australia
Telescope Compact Array using all six antennas. However, baselines
with antenna 6, (which gives the longest baselines - up to 6 km) were
discarded since they produced insignificant correlation amplitudes.
The dual polarization AT receiver was employed with the correlator set
to 256 channels per polarization, and with each channel separated by
15.7 KHz (3.31 km/s).  The observing logs for the three configurations
are given in Table~\ref{t:radlog}.  The integration time was 20 s per
visibility measurement.  Some data in each run were lost to equipment
failure and weather.

The data were reduced using standard software in the {\em AIPS\/}
package (ATNF version).  The relevant tasks used are noted here in
parenthesis.  Bad or suspect data were edited out (TVFLAG, SPFLAG).
Temporal gain and phase drifts were calibrated (CALIB) using the
secondary calibrator, 0906-682 which was observed at 50 min.\
intervals for 3--6 min.  Spectral variations in the calibration were
determined from the observations of 1934-638 and 0407-658 (BPASS).
The absolute flux level was set by the primary calibrator 1934-638
(SETJY, GETJY; flux reference: Walsh, 1992).  The calibration was
checked for the 1.5 km array data set using 0407-658 which is also a
flux standard (Walsh, 1992).  The calibration agreed to within 2\%\ with the
adopted 1934-638 calibration.  The calibrations were applied and the
two polarization channels combined to form Stokes ``I'' data sets
(SPLIT). The continuum was fitted and subtracted in the UV plane
(UVLSF) and the resultant continuum and line data sets were averaged in
time (UVAVG) to one minute per visibility.  The line data sets were
shifted to the heliocentric rest frame (CVEL), and the data from the
three runs were combined (DBCON).  The data were imaged and
``cleaned'' (MX; Schwab, 1984; see also Clark, 1980; H\"ogbom, 1974)
to about the noise level per resultant channel to make both uniform
and natural weighted data cubes (hereafter UN and NA respectively).
For the NA cube, two spectral channels at a time were averaged at the
imaging stage.  The UN data was imaged at full spectral resolution.
The beam sizes at the 50\%\ level are $\Wfo = 45''$ (circular) for the
NA data cube and $\Wfo = 27'' \times 23''$ (${\rm PA} = 0\dg$) for the
UN cube.  The final cubes were made by reconvolving the clean
component images with two-dimensional Gaussian beams having the above
sizes.  Planes of the resultant NA data cube are shown in
Fig.~\ref{f:cube}.

From each cube, maps of total intensity, mean velocity, and line
broadening were constructed from the zeroth, first, and second moments
of the data cubes (w.r.t.\ velocity; MOMNT).  The moment maps and data
cubes were then corrected for the fall-off in primary beam response
with distance from the pointing center (PBCOR).  Table~\ref{t:HIprop}
summarizes some of the properties of the NA and UN data sets including
the beam size, pixel size and noise level in the resultant data cubes.

In order to check the moment analysis results, single Gaussian fits to
the UN data cube were made (XGAUS).  As a further check, and to
examine the line profiles in detail the UN data cube was block
averaged into $30''\times 30''$ pixels, and all the resulting profiles
were examined and fit with a multiple Gaussian fitting algorithm using
the {\em IRAF\/} program SPLOT.

\section{Overview of HI properties}\label{s:morph}

A cursory examination of the \HI\ properties of NGC~2915 shows that
they are very different from its optical
properties. Figure~\ref{f:morph} (plate XXXX) shows three \HI\ surface
density maps, plotted as grey scales.  Panel a shows the NA moment 0
map (45\as\ beam); panel b, the UN moment 0 map ($27'' \times 23''$
beam), and panel c, the XGAUS peak amplitude map.  NGC~2915 clearly
has a spiral morphology, and this is most readily apparent in the NA
map. There appears to be two arms which can be traced back to the ends
of central bar structure.  Figure~\ref{f:na0} shows a contour plot of
the NA moment 0 map, while the UN moment 0 contours are shown in
Fig.~\ref{f:optun0} (Plate XXXX) overlaid on an {\em I\/} band image
obtained with the Anglo-Australian Telescope.  This clearly shows that
the \HI, including the spiral arms, extends well beyond the optical
extent of NGC~2915.

The prominent central bar is about 2.5$'$ (3.9 kpc) long.  It is
most clearly seen in panel~a of Fig.~\ref{f:morph}.  The velocity
field, analyzed in \S\ref{ss:tring}, shows the signature of an oval
distortion demonstrating that this is a real bar, and not a highly
inclined disk. The bar is resolved into two clouds in the UN data as
can be seen in the bottom two panels of Fig.~\ref{f:morph} and
Fig.~\ref{f:optun0}.  Each of these clouds has a neutral gas
mass\footnote{The neutral gas mass is the \HI\ mass multiplied by 1.33
to correct for the neutral He content.} of $\sim 2.5 \times 10^7$
\Msun\ ($1.4 \times 10^7$ \Msun\ if a local background is subtracted)
which is the majority of the mass in the bar.  They are located 17\as\
NW and 36\as\ SE of cluster 1 (paper I).  They encompass many of the
embedded objects discussed in paper I and bracket the brighter central
clusters as well as some of the peculiar structure on the SE side of
the galaxy.  The faint ``jets'', first mentioned by S\'ersic \etal\
(1977) (cf.\ Fig.\ 2 paper I) project outwards from the core of these
clouds. The XGAUS map in panel~c of Fig.~\ref{f:morph} is very noisy,
and clearly shows only the highest S/N features.  In this map the
morphology is reminiscent of a spiral with a bar interior to a
detached ring.

The optically visible portion of NGC~2915 coincides in position,
orientation, and size to the central bar.  Thus while the exponential
optical surface brightness profile (for $R > 35''$) implies  a disk
structure, the \HI\ data suggests that the stars are in a bar
configuration. This interpretation is consistent with the optical
morphology (paper I) which shows enhanced star formation primarily in
the center of the galaxy, but also along the optical major axis,
or the bar ridge line. Such a distribution of star formation is common
in barred galaxies (Friedli \&\ Benz, 1995, and references therein).

There is a large ``hole'' $\sim 130'' \times 180''$ ($3.4 \times 4.6$
kpc) located 310\as\ (8.9 kpc) to the east of the galaxy's center.
Examination of a position-velocity cut through this structure does not
reveal split line profiles, indicating that it is probably not a
kinematic shell, but an interarm clearing.  The \HI\ in NGC~2915 does
not appear to be very porous, as is found to be the case in many
nearby disk galaxies, such as HoII (\markcite{Puche \etal, 1992}).
This may be largely a result of resolution; the HoII data have a
resolution of 66 pc, and typical bubble sizes are a few hundred pc,
while the linear resolution of our data is ten times worse (640 pc).

The velocity field from the NA and UN moment 1 maps are shown in
Figs.~\ref{f:na1} and \ref{f:optun1} (Plate XXXX) respectively.  They
clearly show the expected pattern of a rotating disk.  The disk must
be warped as indicated by the twisting of the kinematic major and
minor axes.  Thus although there are closed velocity contours, it is
not readily apparent whether these are due to a declining rotation
curve, or the warped disk.  The rotation curve is derived in detail in
\S\ref{ss:tring}.  The NA and UN global velocity profiles are shown in
Fig.~\ref{f:globl}.  These were made from the data cubes by summing
the flux in each channel within a ``blanking aperture''.  The aperture
was defined by eye using the total intensity map as a guide.  The twin
horned nature of the profile is the classic profile of a rotating
disk.  The bump near \Vsys\ is due to the large central concentration
of \HI.

Figure~\ref{f:xgb} shows the line of sight velocity dispersion, $b$, map
from the XGAUS profile fitting.  It shows that $b$ increases by several
tens of \kms\ towards the center of NGC~2915. This is a much larger
gradient than typically observed in spiral and irregular galaxies
(Kamphuis, 1992; Lo \etal, 1993).  These maps indicate that pressure
support may be important to the dynamics of NGC~2915.  The line profiles
are discussed in detail in \S\ref{ss:prof}, while the pressure support,
or asymmetric drift, corrections to the rotation curve are discussed in
\S\ref{ss:asymd}.

Table~\ref{t:HIprop} presents some properties measured from our data. 
From the global \HI\ profiles we measured the total line flux $\int S
dV$, the corresponding \HI\ mass, \MHI, the velocity widths at 50\%\
and 20\%\ of the maximum intensity, \Wfo\ and \Wto\ respectively, and
the systemic velocity \Vsys.  That $\int S dV$ from the UN and NA
global profiles agree so well, indicates that the UN data is not
missing much short spacing flux compared to the NA data.  The \HI\
mass, $\MHI = 9.6 \pm 0.2 \times 10^8$~\Msun\ from the NA data, is
significantly higher than that derived by \markcite{Becker \etal\
(1988)}, $\MHI = 7.0 \times 10^8$ (after adjusting to our adopted
distance) using a low resolution \HI\ map derived from Parkes 64m
observations.  We suspect that the flux difference is due to the lower
signal to noise ratio of their observations and perhaps inadequacies
in their beam profile correction.  The systemic velocities measured
from the global profile, \Vsys(global), and the tilted ring analysis,
\Vsys(dynamical) (see \S\ref{ss:tring}), agree well with each other,
and also with the optical velocity \Vsys(optical) $= 468 \pm 5\,\kms$
reported in paper I.

From the total intensity maps we measure $R_{\rm HI}$, the radius
where the face-on \HI\ column density $N(HI) = 5 \times 10^{19}\, {\rm
cm^{-2}}$, and $N_{\rm HI}$(max), the maximum \HI\ face-on column
density.  The correction to face-on is for the mean inclination $i =
59\dg$, as derived below.  The remaining properties in
Table~\ref{t:HIprop} compare the \HI\ and optical properties (paper
I).  \MHI\ is compared to the {\em B\/} band luminosity, $L_B$, and
$R_{\rm HI}$ is compared to the Holmberg radius $R_{\rm Ho}$ and {\em
B\/} band exponential scale length $\alpha_{B}^{-1}$.  The resultant
ratios quantify what is already clear: NGC~2915 is gas-rich and very
extended in \HI.

\section{ISM Dynamics}\label{s:dyn}

\subsection{Tilted ring analysis}\label{ss:tring}

The \HI\ rotation curve was derived using the now standard tilted ring
algorithm, ROTCUR (\markcite{Begeman 1989)}, in a manner similar to
that outlined by \markcite{Martimbeau \etal\ (1994)}.  The UN and NA
first moment maps, and the UN-XGAUS velocity field were analyzed
separately, using ring widths of 25\as, and 45\as\ for the UN and NA
data respectively.  There are up to six parameters for each ring; the
central coordinates $X_c, Y_c$, the systemic velocity \Vsys, the
inclination $i$, the position angle $\phi$ of the kinematic major
axis (receding side), and the rotation velocity,
\Vrot.  The parameters were derived using five iterations of the
algorithm.  In the first iteration all parameters are allowed to be
free.  This iteration gives a feel for how the parameters
vary with radius $R$. In the second iteration, $i$ and $\phi$ are held
fixed while the other parameters are allowed to vary.  The aim of this
iteration is to find the best average value of $X_c, Y_c$, and \Vsys,
which are then held fixed at a constant value for all subsequent
iterations.  The resultant mean \Vsys\ values are reported in
Table~\ref{t:HIprop} as \Vsys(dynamical).  In the third iteration, new
radial profiles of $i, \phi$, and \Vrot\ are derived.  Smooth curves
are drawn through these profiles and the interpolated values of these
parameters are used as the initial guess for these parameters in the
fourth iteration, which is meant to test how stable the solutions are
to small changes in the initial guesses.

It was found that the NA-MOMNT and UN-XGAUS solutions are stable to
small perturbations.  However the inner three rings of the UN-MOMNT
field ($R < 90''$) are not.  No unique solutions could be found.  This
is due to the kinks in the central isovelocity contours seen in
Fig.~\ref{f:optun1}.  Therefore we adopt the UN-XGAUS results for $R \leq
90''$, the UN-MOMNT results for $90'' < R \leq 390''$ and the NA
results for $R > 390''$.  Finally each half (approaching and receding)
of the galaxy was analyzed separately.  This final step allows the
uncertainty of the parameters to be estimated from the level of
asymmetry in the velocity field; the errors in the parameters are taken
to be the absolute difference between the value for the full ring and
either the receding or approaching half, whichever is larger.  The
\Vrot\ errors estimated by this method are always larger than or equal
to the formal errors from the least squares fitting.  The formal
errors in $i$ and $\phi$ are occasionally larger than the asymmetry of
field errors.  In these cases the former is adopted.

The results in terms of $i$, $\phi$ and \Vrot\ for each data set are
shown in Fig.~\ref{f:ring}.  The adopted fit parameters are tabulated
in Table~\ref{t:rot}.  The \HI\ disk shows a strong warp in both
$\phi$ ($\Delta\phi = 40\dg$) and $i$ ($\Delta i = 25\dg$).  The warp
is strongest, especially in $i$, for $R \leq 150''$, with relatively
mild warping ($\Delta \phi = 15\dg$, $\Delta i = 4\dg$) beyond this
radius.  The adopted rotation curve shows a rapid rise to $\Vrot
\approx 80$ \kms\ at $R = 202.5''$. It then remains approximately
constant until rising again for $R > 290''$.  Figure~\ref{f:lv70}
shows a position velocity cut through the NA data cube along the major
axis ($\phi = -70\dg$).  The second rise can be seen directly on both
sides of the galaxy, demonstrating that it is real.

Figure~\ref{f:rresid} shows the residual maps from the ring fits.
Local deviations from the fit have $\vert \Delta V_r \vert \lapeq
15~\kms$.  The residuals in the inner portion show rings of mostly
negative and mostly positive residuals at $R \approx 65''$ and 150\as\
respectively. This indicates that \Vsys\ varies with $R$, suggesting
that the inner portions of the galaxy are not in equilibrium.
Kamphuis (1992) finds \Vsys\ variations in normal spiral galaxies, but
at large radii.  In the center where \Vrot\ is low these deviations
are large enough to cause the isovelocity contour twists and the
problems in the UN-MOMNT fit noted above.  The residuals along the
minor axis have opposing signs up to $\approx \pm 10$\kms\ on either
side of the center ($+$ to the NE, $-$ to the SW).  This is a
signature of an oval distortion\footnote{Since the line widths along
the minor axis are not elevated relative to the major axis, it is not
likely that residuals are due to a galactic wind.}: the kinematic
major and minor axes are non-orthogonal (Bosma, 1981).  The effect can
also be discerned in Figs~\ref{f:na1},\ref{f:optun1}.  The distortion
probably arises from the central bar, and is strongest for the rings
with $200'' \lapeq R \lapeq 450''$.  It shows that the orbits are not
strictly circular for these rings, although one should bear in mind
that the residuals are small compared to the amplitude of $\Vrot
\approx 80$ \kms\ at these radii.

\subsection{Line profiles}\label{ss:prof}

The results of the automatic Gaussian fitting are shown in
Fig.~\ref{f:xgb} which gives a map of $b$ values from XGAUS.  The $b$
map is not symmetric about the center of NGC~2915.  There is
stronger line broadening on the SE side than the NW.  The broadening
is centered on the SE \HI\ cloud (see Fig.~\ref{f:optun0}).  

XGAUS was set to make only single Gaussian fits.  The SPLOT
measurements were designed to test the XGAUS results and measure
complex line profiles.  The position of the splot fits are shown in
Fig.~\ref{f:split}, with the positions of profiles best fit by
multiple Gaussians highlighted. The \HI\ profiles are predominantly
single peaked.  The main region of split lines is confined to within
$2'$ of the center, which is also where the lines become the broadest.
Profiles at the position of both \HI\ clouds are split.  Thus the
total \HI\ width distribution (including splitting) is a bit more
symmetric than the $b$ map shown in Fig.~\ref{f:xgb}.  The average line
split is $\overline{\Delta V_r} = 41$ \kms.   A few representative
profiles and their SPLOT fits are shown in Fig.~\ref{f:examprofs}.

Could the broad profiles be due to beam smearing of the observed steep
rotation curve?  This is not likely, because the knee in the rotation
curve is four to six beam widths from the center, and thus the central
velocity gradient is not that steep. Beam smearing only shows its
effects over one to two beam widths.  We verified this by simulations
of the data including the effects of beam smearing on the UN data cube.  A
change in \Vrot\ of 80 \kms\ in one beam width can result in $b = 27$
\kms.  But we can rule out such a steep rotation curve, because it
would require the knee in the rotation curve to be at $R = 50''$.
Only higher resolution observations can determine if there are large
velocity gradients within any given beam.  However any such gradients
are likely to be turbulent in nature since the broad ($b \geq 20\,
\kms$) line region is eight beam diameters wide.

Figure~\ref{f:compw} compares the SPLOT $b$ measurements with the
azimuthally averaged $b$ from the MOMNT and XGAUS results.  It shows
that the SPLOT and XGAUS results are in good agreement, indicating
that the additional beam smearing in the 30\as\ pixel data has not
substantially increased $b$ with respect to the full resolution UN
data.  The MOMNT analysis, on the other hand, yields $b$ substantially
lower than the Gaussian fits.  This is due to an artificial bias in
the MOMNT analysis technique; MOMNT only uses the data where the
signal is above a certain threshold in a smoothed version of the data
cube.  This systematically excludes the profile wings, and thus the
results are biased towards lower $b$.

The dotted line in Fig.~\ref{f:compw} shows the two pixel resolution of
the UN data cube ($b = 2.8~\kms$). Most profiles are well sampled by
our data, those having $b\lapeq 2.8~\kms$ mostly have low signal to
noise ratios. A notable exception is shown in Fig.~\ref{f:examprofs}f.
At large $R$, the mean $b$ is around 8~\kms, which is typical for disk
galaxies (Kamphuis, 1992).  Figure~\ref{f:examprofs} shows that at this
width the profiles remain well sampled.

\subsection{Asymmetric drift correction}\label{ss:asymd}

The broadening of the \HI\ profiles in the central region is
significant compared to the measured rotation, as can be seen by
comparing Fig.~\ref{f:compw} with Fig.~\ref{f:ring}. The random
motions of the gas may then provide significant dynamical support in
the central region.  It is not clear that the gas there is in
dynamical equilibrium.  Indeed the asymmetry of the $b$ map suggests
that it is not.  Nevertheless, here we will assume that gas is in
equilibrium and account for the pressure support by estimating an
``asymmetric drift'' correction following Oort (1965; see his eq.\
10).  We assume that the gas velocity ellipsoid (\ie\ pressure) is
isotropic.  This greatly simplifies the correction\footnote{Binney and
Tremaine (1978) treat the case of the asymmetric drift in collisionless
systems (stars only) which have non-isotropic velocity ellipsoids.}.
In a cylindrical coordinate system, the radial ($R$) acceleration
$K_R$ about the rotation axis $z$ is given by
\[ -K_R = \frac{\Vrot^2}{R}
       - b^2 \frac{\partial( \ln [\rho b^2])}{\partial R}, 
\] 
where $\rho$ is the density.  If there were no pressure support the
orbits would be circular with velocities $V_c$ given by
\[ \frac{V_c^2}{R} = -K_R. \]
The density can be written as $\rho = \Sigma_g/(2h_z)$, where $h_z$ is
the vertical scale height and $\Sigma_g$ the gas surface density (here
taken as \HI\ plus neutral He).  The circular velocity is then 
\begin{equation}  
V_c^2 = \Vrot^2 + \sigma_D^2 \label{e:ad1}
\end{equation} 
where $\sigma_D$ is the asymmetric drift correction given by
\begin{equation} 
\sigma_D^2 = - Rb^2 \left[ \frac{\partial \ln(\Sigma_g)}{\partial R} 
               + 2\frac{\partial \ln(b)}{\partial R} -
               \frac{\partial \ln(h_z)}{\partial R} \right].\label{e:ad2}
\end{equation} 
Usually the contributions of each of the gradients are zero
or negative, making the correction positive.  We do not have a direct
measurement of $h_z$ but will assume ${\partial \ln(h_z)} / {\partial
R} = 0$.  The other two terms can be estimated from the radial
profiles of $\ln(b)$ and $\ln(\Sigma_g)$ which are shown in
Fig.~\ref{f:asymd}. Both profiles have a core-halo structure and 
are well fit by a function consisting of a
Gaussian centered at $R = 0$ on top of a polynomial (in $R$)
background:
\begin{eqnarray}
\ln(b) & = & 2.40 + 6.57\times 10^{-4}R \nonumber \\
       &   & + 1.23\exp(-\{R/85.1\}^2/2) \\ \label{e:bfit}
\ln(\Sigma_g) & = & 0.812 + 1.057\times 10^{-3}R \nonumber \\
              &   & - 7.13\times 10^{-6}R^2 \nonumber \\
             &   & + 1.51\exp(-\{R/56.2\}^2/2) 
\end{eqnarray}
where $R$ is in arcsec, $b$ is in \kms, and $\Sigma_g$ is the gas
surface density in $\Msun\, {\rm pc^{-2}}$.  The fits were unweighted
and have an rms of 0.05 and 0.07 in $\ln(b)$ and $\ln(\Sigma_g)$
respectively.  These fits (converted to linear units) are shown in
Fig.~\ref{f:asymd}, as well as the $\sigma_D$ curve resulting from
differentiating the fits.  The last two columns of Table~\ref{t:rot}
tabulate $\sigma_D$ and the corrected circular velocities, $V_c$.  As
expected the $\sigma_D$ corrections are quite significant for $R \leq
152''$.  We can now proceed to model the mass distribution, with the
caveat that the
\HI\ may not be equilibrium at these radii.

\subsection{Mass models}\label{ss:mm}

Three component mass models were fit to the rotation curve. The
components are the stellar disk, a neutral ISM (\HI) disk, and a dark
matter (DM) halo.  The stellar mass distribution is determined from
the surface brightness profiles in paper I, and has one free parameter
${\cal M}/L_B$.  The mass distribution of the \HI\ disk (ncluding
neutral He) is determined from the observations presented here, and
there are no free parameters.  The DM halo is assumed to have a
spherically symmetric density distribution given by
\begin{equation} 
\rho = \frac{\rho_0}{1 + (R/R_c)^\gamma} \label{e:mm}
\end{equation}
where the free parameters are the central density $\rho_0$ and the core
radius $R_c$.  Here we adopt $\gamma = 2$.  For this density
distribution the rotational velocity at large $R$, $V_{\infty}$, and halo
velocity dispersion $\sigma_0$ are given by (Lake \etal\ 1990):
\[ V_{\infty}^2 = 4 \pi G \rho_0 R_c^2 = 4.9 \sigma_0. \]
Fitting was done with a $\chi^2$ minimization technique.  Thus the
fitted parameters depend not only on the input data, but also the
associated errors.

We experimented with numerous schemes for assigning errors to the
$V_c$ values in Table~\ref{t:rot}. These included adopting the \Vrot\
errors, combining the \Vrot\ errors with likely $\sigma_D$ errors, and
taking local averages of the errors over several rings.  In the end we
adopt a uniform error in $V_c$ of 3 \kms.  This corresponds to the
average error in \Vrot\ for $R > 90''$.  The other methods invariably
assign less weight to the inner points than to the outer points.  The
$\chi^2$ minimization then effectively sacrifices the fit in the inner
regions in order to make a marginal improvement at large $R$.  Only
equal weighting produces a reasonable fit at all $R$.  So for the
adopted fits we are no longer doing true $\chi^2$ minimization, but
effectively a non-weighted least squares minimization.  Four models
are discussed.  They are illustrated in Fig.~\ref{f:mm}, and
their parameters are tabulated in Table~\ref{t:mm}.  Each panel of
Fig.~\ref{f:mm} shows the fitted rotation curve and the contribution
of each of the components.

Model A is the best fit to the $V_c$ data.  All three parameters are
allowed to be free, with the only constraint that they be $\geq 0$.  The
resulting fit has ${\cal M}/L_B = 0.0$, i.e.\  the disk is not needed to fit
the rotation curve.  The halo parameters are for a very compact and
dense core, the most extreme of the models generated here.  

Model B is a fit to \Vrot\ instead of $V_c$; in effect it ignores the
pressure support.  The difference between Models A and B is thus
indicative of the uncertain dynamical state of the central region.
Neglecting the asymmetric drift correction results in a larger, less
dense core.  The effect on the inferred core parameters is quite
dramatic, a factor of about 1.5 in $R_c$ and more than 2 in
$\rho_0$. Since we may have underestimated $\sigma_D$, it is possible
that NGC~2915's halo is even more dense and compact than that of Model
A.  Like model A, model B yields ${\cal M}/L_B = 0.0$.

Model C shows a fit where the DM halo is a non-singular isothermal
sphere (e.g.\ Binney \&\ Tremaine 1987).  For this density distribution
\[ \sigma_0 = \frac{2}{3} R_c \sqrt{\pi G \rho_0} 
= \frac{V_\infty}{\sqrt{2}}. \] This is the form of the DM
distribution usually favored by Carignan and collaborators (e.g.\
Puche \&\ Carignan, 1991, and references therein).  It is clear that
model A is a better fit over all radii, but model C produces as good
or better of a fit for $R \leq 6$ kpc.  The difference between Models
A and C illustrates the uncertainty due to which arbitrary form of the
halo distribution is adopted.  The isothermal sphere and the analytic
form of eq.~\ref{e:mm} have significantly different scaled density
profiles over their first few $R_c$.  Thus model C has very different
parameter values than A and B.  It has a much less dense core, and 
it is the only one of the three with a non-zero ${\cal M}/L_B$ as 
the best fit.

Models A-C give the best fit value with all parameters free.  The
resulting limits on ${\cal M}/L_B$ are lower than one would expect for
the red color of the outlying diffuse stellar population in NGC~2915.
For Model D we return to the analytic halo models given by
eq.~\ref{e:mm} and fix ${\cal M}/L_B = 1.2$, which represents
approximately the maximum the disk component can contribute to match
the innermost $V_c$ data point.  This mass to light ratio is in the
range expected for the colors of NGC~2915 ($1 \lapeq M/L_B \lapeq 3$;
Bruzual and Charlot, 1993). The resultant halo parameters turn out to
be identical to model B.  In this case the disk contribution to
$V_c(R)$ is of the same order as the $\sigma_D$ corrections.  Likewise
models of type C, but with no $\sigma_D$ corrections, produce
negligible disk contributions.  Thus the strength of the disk
component critically depends on the $\sigma_D$ corrections.  Fits
using the {\em R\/} band light profile (paper I) are not significantly
better in fitting the rotation curve than the {\em B\/} profile used
here.  Model D is our preferred fit, since the limits on ${\cal
M}/L_B$ are implausibly low for models A and B, and model D is
clearly a better fit than model C.

Note that models A, B and D have very similar values $V_\infty$ with
low uncertainties.  This is because $V_\infty$ essentially determines
the level of the flat part of the rotation curve, which is constrained
much better than the inner portions.  $V_\infty$ is lower for model
C because the isothermal sphere reaches a maximum $V_c$ of $1.1
V_\infty$, and does not approach $V_\infty$ until $R \gapeq 10 R_c$
(Binney \&\ Tremaine 1987).

None of the models can produce the exact shape of the rotation curve,
especially the upturn in \Vrot\ at large $R$.  We performed fits where
$\gamma$ in eq.~\ref{e:mm} is a free parameter.  These yield $\gamma =
1.93$ and are not a significant improvement over Model A. The $\pm 8$
\kms\ modulations in the $V_c$ profile occur where the $b$ profile and
residual maps show significant non-circular motions.  One should not
over-interpret them as necessarily reflecting structure in the mass
distribution.

\section{Discussion}\label{s:disc}

\subsection{Comparison with other galaxies}\label{ss:compare}

NGC~2915 is an intriguing galaxy because its optical properties are
those of a weak BCD (paper I) while its \HI\ morphology and global
profiles are those of an Sd-Sm disk galaxy (cf.\ Roberts and Haynes,
1994; Shostak, 1978).  It has both a bar and a strong warp.  These are
normal properties of spiral galaxies (Holmberg 1958; Bosma 1991).
NGC~2915 appears to be a two armed system.  The arm
structure is not flocculent like the transient features seen in
simulations of shearing disks (e.g.\ Gerola \&\ Seiden, 1978),
despite occurring in the flat (shearing) part of the rotation curve.
The arms are therefore probably due to a spiral density wave. Although
the kinematic signature of such a wave is not apparent, this may be
due to the quality of the data (low signal to noise, insufficient
resolution).  

NGC~2915's \HI\ extends well beyond $R_{\rm Ho}$ which is common for
galaxies with high ${\cal M}_{\rm HI}/L_B$ ratio (Huchtmeier \&\
Seiradakis, 1985; Hoffman \etal, 1993; van Zee \etal\ 1995).  Its
$\Sigma_g$ profile has a core-halo structure.  This is less
common for spiral galaxies which typically exhibit an \HI\ deficit or
plateau in the center (Wevers \etal\ 1986; Warmels, 1988), although
similar cores have been found in irregular galaxies (e.g.\ NGC~55;
Puche \etal, 1991).  NGC~2915's velocity dispersion profile also
stands out for the high value, $b\approx 40$ \kms, in the center.
This is much higher than seen in normal (non-bursting) disk galaxies,
which typically reach a central maximum of $b \approx 15$ \kms,
dropping to 6 to 8 \kms\ at large radii (e.g.\ Dickey \etal, 1990; see
also comparisons of Kamphuis, 1992; Lo \etal\ 1993).  Note that
although NGC~2915 has a normal $b= 8$ \kms\ beyond the optical radius,
it is unclear how it and other galaxies with extended \HI\ disks
maintain this velocity dispersion beyond any apparent heat sources.

The \HI\ properties of NGC~2915 are less abnormal compared to the
relatively few BCD and amorphous galaxies imaged in \HI.  These
galaxies typically contain extended \HI\ disks well beyond the optical
radius, core-halo $\Sigma_{\rm HI}$ profiles and $b$ profiles reaching
25 to 40 \kms\ in the center (Taylor \etal\ 1994; 1995; Hunter \etal\
1994; Meurer 1994; Brinks 1988; Viallefond \&\ Thuan, 1983).  The
properties of NGC~2915 are also reminiscent of the starburst IBm
galaxy NGC~4449 which has an \HI\ disk over ten times larger than its
optical size (Bajaja \etal, 1994).  This is not to say that NGC~2915
is a typical BCD.  None of the BCDs that have been imaged at \HI\ have
spiral arms.  Only 8\%\ of BCDs have double horn profiles and the
large gas to stars ratio $M_{\rm HI}/L_B = 2.7$ of NGC~2915 also
stands out amongst BCDs, which have a median $M_{\rm HI}/L_B = 0.2$
(Thuan \&\ Martin, 1981).

Less is known about the DM content of BCDs. This paper presents the
first mass model decompositions of a BCD.  When \HI\ extends well
beyond $R_{\rm Ho}$ and the rotation curve is flat, the result is a
high ${\cal M}_T/L_B$ ratio.  This is certainly true for NGC~2915
which has ${\cal M}_T/L_B = 76$ (model D) compared to DDO154 with
${\cal M}_T/L_B = 74$ (Carignan and Beaulieu, 1989; Hoffman
\etal\ 1993) and DDO170 with ${\cal M}_T/L_B = 57$ (Begeman \etal\
1991; Lake \etal\ 1990).  Thus NGC~2915 is one of the darkest disk
galaxies known.  We note that the ${\cal M_T}/L_B$ values for giant
ellipticals and dwarf spheroidals are similarly large (e.g.\ Grillmair
\etal, 1994; Mateo \etal\ 1993).  

It is not just the extent of NGC~2915's \HI\ that distinguishes its
remarkable DM properties.  The DM core also appears to be
extraordinarily dense with $\rho_0 \approx 0.1\, \Msun\, {\rm pc}^{-3}$
which is about an order of magnitude denser than typically found in disk
galaxies (Skillman \etal\ 1987; Puche \&\ Carignan, 1991; Begeman
\etal\ 1991), but in the range of that found in dwarf spheroidal
galaxies $\rho_0 = 0.07$ to 1.5 $\Msun\, {\rm pc^{-3}}$
(\markcite{Mateo \etal, 1993}).

Neither the core size, $R_c \approx 1$ to 3 kpc, nor $V_\infty = 73$ to
90 \kms\ (depending on model form) are particularly distinguishing;
similar values of $R_c$ are found in dwarf irregular galaxies, while low
to moderate mass spirals have similar values of $V_\infty$ (Puche \&\
Carignan 1991; and references therein).  But having both a high 
$V_\infty$ and short $R_c$ is rare.  One exception is NGC~5585. C\^ot\'e
\etal\ (1991) fit a mass model, including an isothermal DM halo, to its
rotation curve and find $R_c = 2.8$ kpc, $V_\infty = 76$ \kms, and
$\rho_0 = 0.060\, \Msun\, {\rm pc^{-3}}$, i.e.\ nearly identical to our
model C. But, whereas NGC~2915 is a BCD, NGC~5585 is a Sd IV spiral with
$\RHo = 6.5$~kpc, over twice as large at optical wavelengths as
NGC~2915.  This striking difference suggests that the optical morphology
of galaxies is independent of the properties of their DM halo. 
NGC~5907, with $\rho_0 \approx 0.08 - 1.9\, \Msun\, {\rm pc^{-3}}$
(\markcite{Sackett \etal, 1994}) also has one of the highest inferred
$\rho_0$ for a disk galaxy.  But in this case $R_c$ is not well
constrained by the rotation curve and $\rho_0$ refers to faint luminous
matter, and not strictly speaking to, ``dark'' matter.

It should be noted that most of the studies quoted above adopt
distances based on $H_0 = 75\, {\rm km\, s^{-1}\, Mpc^{-1}}$. Although
the distance to NGC 2915 does not explicitly depend on $H_0$, it is
tied to the distance to NGC~5253 derived by \markcite{Sandage \etal\
(1994)} who find $H_0 = 54\, {\rm km\, s^{-1}\, Mpc^{-1}}$ (paper I).
We reran our preferred model, D, taking $D= 3.1$~Mpc for NGC~2915, which
is more appropriate for $H_0 = 75\, {\rm km\, s^{-1}\, Mpc^{-1}}$
(here $D$ is derived from $V_r$ using a Virgo-centric flow model, and
the appropriate $D_{\rm Virgo}$), and fixing ${\cal M}/L_B = 2.0$.  The
results make NGC~2915 appear more extreme: $R_c = 0.71 \pm 0.08$ kpc,
and $\rho_0 = 0.31 \pm 0.06\, \Msun\, {\rm pc^{-3}}$ ($\sigma = 30.5
\pm 0.7$ \kms).  If NGC~2915 were this close it would have ${\cal M}_T/L_B =
130$ at its last measured point.  A comparison of optical and DM halo
properties of galaxies employing a uniform $H_0$ would be highly
desirable.

\subsection{Central Energetics}\label{ss:core}

The neutral ISM is very energetic in the central bar area of NGC~2915.
The energy is displayed in two forms: that in expanding structures,
and general turbulence. The former is characterized by split lines
implying an expansion velocity of $V_{\rm exp} = 0.5\Delta V$, where
$\Delta V$ is the amount of line splitting.  The expansion energy is
\[ E_k({\rm exp\/}) = \frac{1}{8} {\cal M}_g \Delta V^2; \]
we take the mass involved in the expansion to be ${\cal M}_g = 10^8\,
\Msun$ (the neutral gas mass within the central bar area) and
$\Delta V = 40$ \kms, and derive $E_k({\rm exp\/}) = 4\times 10^{53}\, {\rm
ergs}$. This is an over-estimate since not all of ${\cal M}_g$ is likely to
be in expanding structures.  For an isotropic velocity distribution
the turbulent energy is given by
\[ E_k({\rm turb\/}) = \frac{3}{2}{\cal M}_g\langle b_{HI} \rangle^2, \]
where $\langle b_{HI} \rangle = 35$ \kms\ is the mean velocity
dispersion in the core.  Thus $E_k({\rm turb}) =
3.8\times 10^{54}\, {\rm erg}$.  The energy in turbulence appears to
be significantly greater than that in expansion, although beam
smearing may cause some confusion between bulk flows and turbulence.
We adopt a total kinetic energy of
\[ E_k = E_k({\rm exp}) + E_k({\rm turb}) = 4.2 \times 10^{54}\, {\rm erg}.\]

What is the source for this kinetic energy?  Figure~\ref{f:optun2}
(plate XXXX) shows the $b$ contours of Fig.~\ref{f:xgb} overlaid on
the \Halpha\ image of Marlowe \etal\ (1995).  This shows the ``smoking
gun'' that implicates hot young stellar populations with the energized
neutral ISM.  There are two \Halpha\ bubbles, of a blister morphology.
That is their ionizing clusters are at the edges of the bubbles (near
the galactic center), and the bubbles are expanding away from these
clusters.  While, the profiles are broad over the whole ionized
region, notice that The protrusion in the $b = 30$ \kms\ contour about
25\as\ to the north of the optical center clearly corresponds to the
\Halpha\ bubble extending along the optical minor axis. The most
energetic \HI\ is associated with the SE \HI\ cloud.  It is at the
edge of the SE \Halpha\ bubble which appears to be expand into it.
Thus the broad \HI\ profiles are associated with the turbulent ionized
ISM and in particular the expanding \Halpha\ bubbles.  The ultimate
energy source is the stellar winds and SNe produced by the young
central stellar populations.

For a solar metallicity population with a Salpeter (1955) IMF and a
constant star formation rate, the equilibrium ratio $L_{\rm
mech}/L_{\rm H\alpha} = 1.8$ (Leitherer and Heckman, 1995), where
$L_{\rm mech}$ is the rate mechanical energy is produced by supernovae
explosions and stellar winds.  From the $L_{\rm H\alpha} = 6.4 \times
10^{39}\, {\rm erg\, s^{-1}}$ (paper I; Marlowe \etal\ 1994) we deduce
$L_{\rm mech} \approx 1.1 \times 10^{40}\, {\rm erg\, s^{-1}}$.  The
crossing time of the core is $t_{\rm cross} \approx 3.1\, {\rm kpc} /
35\, \kms = 87$ Myr (i.e.\ the length of the region divided by the
typical velocity).  During this time the mechanical energy release of
the central population is $E_{\rm mech} = L_{\rm mech} t_{\rm cross} =
3 \times 10^{55}$ erg, or seven times $E_k$.  Only 10\%\ -- 20\%\ of
$L_{\rm mech}$ is likely to end up as kinetic energy, the rest being
thermalized (Ostriker \&\ McKee, 1988; Mac Low \&\ McCray, 1988). So
to a first approximation, the energy released by the young central
population in NGC~2915 is sufficient to explain the observed kinetic
energy in the neutral ISM.

\subsection{Tully-Fisher Relationship}\label{ss:tfr}

Figure \ref{f:tfr} shows NGC~2915 and DDO154 in the absolute magnitude versus
\HI\ line width diagram.  The tight correlation shown by the filled
circles (Sandage and Tamman, 1976) and $\times$'s (Sandage, 1988), is
the Tully-Fisher (1977) relationship (TFR) for nearby
galaxies. NGC~2915, like DDO154 (Carignan and Beaulieu, 1989), falls
way off this correlation.  This discrepancy is also apparent in other
recent calibrations of the TFR (e.g.\ Pierce and Tully, 1988) when the
data is adjusted to the same value of $H_0$.  Carignan and Beaulieu
(1989) note that DDO154 shows the TFR may break down at the low mass
end.  NGC~2915 shows that it does not always hold even for moderate
masses.

Milgrom and Braun (1988) note that the TFR should probably be considered
a correlation between luminous matter - defined as stars and all phases
of the ISM - and line width, not just optical luminosity and line width.
The TFR holds for most galaxies because the stellar mass dominates over
the ISM.  However, for NGC~2915 and DDO154 the neutral gas component is
more significant than the stellar component.  If their ISM were
converted to stars with a ${\cal M}/L_B = 0.3 - 0.4\, (M/L_B)_\odot$
(higher if there is a significant mass in molecular gas) they would fall
on the TFR.  Thus under this scenario, the discrepancy with the TFR may
indicate that the disks of DDO154 and NGC~2915 are relatively unevolved
compared to normal galaxies.  

There still remains the problem of the distribution of luminous matter
and the TFR.  In a normal TFR galaxy, the stellar distribution, which
dominates the luminous mass, is much more concentrated than the DM.
The contributions of each component to the rotation curve produces
similar maximum values $V_{\rm max}$, but at different radii, such
that the overall rotation curve is nearly flat at $V_{\rm rot} =
V_{\rm max}$ from about a disk scale length out to the last measured
point.  This is the so-called ``disk-halo conspiracy''.  If the
neutral ISM of the discrepant galaxies were transformed to stars {\em
in situ\/} such that the resultant luminosity is on the TFR, the
stellar distribution will be coextensive with the DM (see below) and
have $V_{\rm max}$ well below that of the DM.  In other words there
would be no disk-halo conspiracy in such cases.  Thus NGC~2915 and
DDO154 would not look like normal TFR galaxies even if their ISM were
converted into stars, unless their ISM were to become centrally
concentrated before star formation.  Likewise, one could adjust the
distance or dust extinction $A_B$ (see paper I) such that NGC~2915
falls on the TFR, but there would still be no disk-halo conspiracy and
the stellar ${\cal M}/L_B$ would be negligible.  So even then NGC~2915
would be unlike any normal TFR galaxy.

The coincidence of the central $b = 40\, \kms = \sigma_0$ of the DM
halo is perhaps a new twist on the disk-halo conspiracy (although one
case does not make a conspiracy).  We speculate on the meaning of this
coincidence in \S\ref{s:conc}.

\subsection{Star Formation Threshold}\label{ss:thresh}

Figure~\ref{f:surfden}a shows the surface density of the contributors
to the mass of NGC~2915 from model D: the neutral gas, $\Sigma_g$, the
stars, $\Sigma_*$, and DM halo, $\Sigma_{\rm DM}$.
Figure~\ref{f:surfden}b shows the ratio of the first two of these with
$\Sigma_{\rm DM}$.  It is immediately apparent that the stars do not
trace the DM very well, but the neutral gas does.  A relatively flat
$\Sigma_g/\Sigma_{\rm DM}$ profile is commonly observed in disk
galaxies, as first pointed out by Bosma (1981; more recent examples
include Carignan \&\ Beaulieu, 1989; Carignan \etal\ 1990).  Rotation
curves of some galaxies are well fit by scaling the $\Sigma_g$ profile
by a constant factor to obtain the DM contribution (Carignan \&\
Beaulieu 1989; Broeils, 1992).

Figure~\ref{f:surfden} also shows the critical density for high mass
star formation, $\Sigma_{\rm crit}$ (Kennicutt, 1989) and $\Sigma_{\rm
crit}/\Sigma_{\rm DM}$. $\Sigma_g$ is less than $\Sigma_{\rm crit}$ by
a factor of about 3 (range of 2 to 9), even in the center of NGC~2915
where high mass star formation is seen.  When only the neutral ISM is
considered, NGC~2915 does not obey Kennicutt's star
formation law. The discrepancy between $\Sigma_g$ and $\Sigma_{\rm
crit}$ is worst in the center of NGC~2915 because $\Sigma_{\rm crit}
\propto b$, and this is where $b$ is the highest.  However, even
adopting a constant $b = 5$ \kms\ profile does not bring $\Sigma_{\rm
crit} < \Sigma_g$.  This is contrary to what is found for a wide
variety of galaxies including both low surface brightness galaxies
(van der Hulst \etal\ 1993) and other BCDs (Taylor \etal, 1994) which
obey the Kennicutt law when {\em only\/} the neutral ISM is
considered. Of course there may be no real discrepancy, since much of
the ISM may be in molecular form. Star formation in NGC~2915 occurs at
$R\leq 30''$ (paper I), where $\Sigma_{\rm crit}/\Sigma_g \approx 9$
and $M_g(R \leq 30'') = 3.1 \times 10^7$ \Msun.  Thus to satisfy
Kennicutt's law there should be $M_{\rm molecular} \geq 3 \times 10^8$
\Msun.  Only about 1\%\ of the so-called dark matter need be molecular
gas for the Kennicutt law to hold.

The peak \HI\ surface density occurs in the two central \HI\ clouds and
is $\Sigma_{\rm HI} = 2.4 \times 10^{21}\, {\rm cm^{-2}}$ or after
inclination correction $\Sigma_{\rm HI} = 1.2 \times 10^{21}$ 
cm$^{-2}$.  Recent high mass star formation is seen near the positions
of both these clouds.  Star formation in NGC~2915 is consistent
with ``simple'' \HI\ threshold schemes for high mass star formation; for
example Gallagher and Hunter (1984) propose that efficient high mass
star formation occurs where $\Sigma_{\rm HI} >  \Sigma_{\rm HI,crit} = 5
\times 10^{20}\, {\rm cm^{-2}}$, while Skillman (1987) similarly
proposes that $\Sigma_{\rm HI,crit} = 10^{21}\, {\rm cm^{-2}}$.
Thus although, the more complex star formation law of Kennicutt
(1989) does not hold for NGC~2915, a simple threshold law does.

Kenney \etal\ (1993) show how tidal shear can limit star formation via
cloud destruction.  In NGC~2915 and the BCDs in the sample of Taylor
\etal\ (1994), the star forming region and indeed most of the optically
visible galaxy is located in the rising portion of the rotation curve. 
This is where the rotation is almost solid body, and thus tidal shear is
minimized.  Similarly the galaxies in the Skillman (1987) study all have
solid body rotation where their \HII\ regions are located.

\section{Conclusions / Summary}\label{s:conc}

Our detailed \HI\ study of the weak BCD galaxy NGC~2915 shows it to have
a nature unexpected from its optical appearance.  The \HI\ morphology
and global properties (mass, line width) are those of a late type barred
spiral galaxy.  The presence of a bar is confirmed by the signature of
an oval distortion in the velocity field.  At our highest resolution the
bar is resolved into two \HI\ clouds separated by 1.5 kpc.  The \HI\
line of sight velocity dispersion in this central region is very high,
$b \approx 40$ \kms, and the \HI\ line is split by $\Delta V \approx 40$
\kms\ at the position of both \HI\ clouds.  The energetics of the
central region can be accounted for by the known young stellar
populations there.  The \HI\ extends well beyond the optical galaxy,
out to five times the Holmberg radius.  At these radii, the dynamics
are dominated by rotation, although there are indications of an oval
distortion in the velocity field, and some warping of the disk.  The
rotation curve is approximately flat for $R > 4$ kpc, but slightly
rising at the last measured points ($R \gapeq 10$ kpc).  For $R < 3$
kpc, random motions are significant and there are indications that
the systemic velocity varies with radius.  It is not clear whether the
system is in equilibrium at these radii.

The rotation curve can not be accounted for by the luminous components
alone.  Large amounts of dark matter (DM) are required, dominating at
nearly all radii.  Within the radius of the last measured point ($R =
15.2$ kpc) the total mass to blue light ratio is ${\cal M}_T/L_B = 76$,
making NGC~2915 one of the darkest disk galaxies known.  The complex
dynamics of the central region, make interpretation of mass model fits
to the rotation curve difficult.  Despite this uncertainty,
it is clear that NGC~2915's DM distribution has an unusually dense (central
density $\rho_0 \approx 0.1\, \Msun\, {\rm pc^{-3}}$) and compact core
(radius $R_c =$ 0.8 to 2.5 kpc).  The disk ${\cal M}/L$ ratio is not
well constrained by our models, although it seems likely that the
stellar component is less massive than the neutral ISM disk.

NGC~2915 does not obey the Tully-Fisher (1977) relationship, being under
luminous by a factor of ten for its line width (or rotational
velocity).  Nor does it obey the Kennicutt (1989) star formation
threshold law.  Unless there is considerable amount of molecular
material, its gas density is always below the critical density for high
mass star formation, even in the center where high mass stars are
observed to have formed recently.  However, it does obey simple critical
density threshold models for high mass star formation (c.f.\ Gallagher
\&\ Hunter 1984; Skillman 1987). 

Finally we speculate on the nature of star formation in NGC~2915, a
topic our observations shed new light on.  In particular they reveal
two key ingredients in its recipe for regulating star formation: a
bar, and a deep central potential.  Friedli \&\ Benz (1995) showed
through numerical simulations that there is a complex interaction
between star formation, the ISM, and the bar in isolated barred
galaxies.  The bar redistributes angular momentum resulting in an
inflow along the bar.  Some star formation along the bar energizes the
ISM (via the resultant supernovae explosions and stellar winds) enough
to inhibit additional star formation until the material reaches the
center.  The coincidence in velocity dispersions, that of \HI\
measured in the center of the galaxy, and that inferred for the DM
halo, suggests that the intensity of the central star formation is
regulated by the depth of the potential well (Bothun \etal, 1986); if
the star formation becomes too strong the ISM is thrown clear of the
core and star formation is halted, at least temporarily.  Some models
indicate that such an outflow can contribute to the spin down of the
disk resulting in further inflow (Corbelli \&\ Salpeter, 1988;
Charlton \&\ Salpeter, 1989) thus establishing a ``galactic
fountain''.  The stellar component of NGC~2915 and other BCD/amorphous
galaxies is limited to within the turnover radius indicating that
tidal shear probably strongly inhibits star formation. The optical
images of NGC~2915 has the size and orientation of the central bar,
suggesting that the stellar counterpart is a bar, or at least that the
bar axis is coincidentaly the optical major axis. A scenario of bar and
DM regulated star formation for NGC~2915 does not require interaction,
which is convenient since NGC~2915 is in a low density environment.
However, there is one possible interaction partner (Appendix) so an
interaction model such as that sketched by Noguchi (1988) or Taylor
\etal\ (1993) can not be ruled out.

\acknowledgements

We thank Amanda Marlowe and Tim Heckman for making \Halpha\ images and
\'Echelle spectra available, and Chris Lidman who assisted in
obtaining and reducing {\em I\/} band images of NGC~2915 and \skiggy.
Stimulating discussions with Albert Bosma, Annette Ferguson, Chris
Mihos, and Renzo Sancisi greatly influenced this paper.  Critiques of
an early draft of this paper from Rosie Wyse, Tim Heckman, Michael
Dahlem, Ron Allen, and Stefano Casertano improved its content and
presentation.  We are also grateful to the anonymous referee for
comments that helped improve this paper.  Luc Turbide and Harry Payne
provided essential software support.  GRM was employed at Universit\'e
de Montr\'eal, and The Space Telescope Science Institute during the
early phases of this investigation.  We are grateful to the directors
of the Australia Telescope National Facility and the Space Telescope
Science Institute for the use of their excellent facilities.  We thank
the Narrabri staff of the ATNF for maintaining their superb facility,
and efficiently bringing the Compact Array back on line after
thunderstorms.  Literature searches were performed using NED, the
NASA/IPAC Extragalactic Database, a facility operated by the Jet
Propulsion Laboratory, Caltech, under contract with the National
Aeronautics and Space Administration.  GRM gratefully acknowledges
funding through NASA grant NAGW-3138, administered by The Johns
Hopkins University.

\appendix

\section{Appendix: an interaction partner?}\label{ss:skiggy}

Taylor \etal\ (1993; 1995; 1996) surveyed apparently isolated BCDs for
companions with the rationale that interaction with an unseen
companion may be responsible for the BCDs present burst of star
formation.  Such companions were successfully found in 12/21 
galaxies in their sample.  We unsuccessfully searched our
\HI\ data cube for faint companions.  However the primary beam of the
AT dishes is $\Wfo = 33'$ which corresponds to a radius of 25 kpc. At
a typical relative velocity of 300~\kms\ an interacting partner can
transverse this distance in less than 100~Myr.

We searched the NASA/IPAC Extragalactic Database (NED)\footnote{The
NASA/IPAC Extragalactic Database (NED) is operated by the Jet
Propulsion Laboratory, California Institute of Technology, under
contract with the National Aeronautics and Space Administration.} for
possible interaction partners within 5\dg\ of NGC~2915.  There are no
galaxies with cataloged velocities within 500~\kms\ of NGC~2915's.
However there is one low surface brightness object, \skiggy, which is
separated from NGC~2915 by 42$'$, a projected distance of 64~kpc, and
has no measured velocity.  Is \skiggy\ the interacting partner of
NGC~2915?

Unfortunately the answer is not yet clear.  Corwin \etal, (1985) note
that it is not certain whether \skiggy\ is a low surface brightness
galaxy or diffuse nebulosity associated with our galaxy.
Figure~\ref{f:skiggyi} shows an {\em I\/} band CCD image of
\skiggy\ obtained using the F1 imaging system on the AAT.  It was made
from four 200s exposures obtained during non-photometric
conditions. There are numerous foreground stars over the face of
\skiggy, but no obvious overabundance of sources that would suggest
\skiggy\ is resolved.  The main body has a diameter of $\sim 130''$.
There are faint extensions to the SE and NW yielding a total diameter
of $\sim 270''$.  The overall symmetry of the system suggests that it
is a galaxy.

An \HI\ spectrum in the direction of \skiggy\ was obtained with the
Parkes 64m radio telescope. The observations cover the range $V_r =
-1000$ to $+4000$ \kms\ split over 1024 channels and reach a noise
level of 50 mJy.  There are no obvious \HI\ features in the spectrum
except for strong galactic \HI\ and a narrow spike having $V_r =
300$~\kms, $W_{50} = 30$ \kms, and $\int S dv = 0.72 \pm 0.06$ Jy
\kms. This probably corresponds to the edge of the high velocity cloud
HVC 300--24+274, which \markcite{Morras (1982)} shows to have narrow
profiles and velocities varying over the range $240 \lapeq V_r \lapeq
330$ \kms.

We conclude that there is no evidence that \skiggy\ is interacting with
NGC~2915. However, it can not yet be ruled out as an interaction
partner.  For an assumed 50~\kms\ line width the \HI\ spectrum should be
able to detect $\MHI > 2.6 \times 10^6$ \Msun\ with $S/N > 5$ at
NGC~2915's distance.  If \skiggy\ is at the distance of NGC~2915 it is
gas poor.


\clearpage

\pino

Fig.\ \ref{f:cube}. Contour plots of the channels with
signal in the CLEANed NA data cube. The contour levels are at 10, 25,
50, 75, and 90 percent of the peak surface brightness of 4.7 Jy
beam$^{-1}$  ($N_{\rm HI} = 1.69 \times 10^{20}\, {\rm cm^{-2}}$). The
$V_r$ in \kms\ of each channel is shown in the upper right of each
panel.  The beam size ($\Wfo = 45''$) is shown on the lower left of
the upper left panel, and again four panels beneath it.

Fig.\ \ref{f:morph}. Grey scale representations of \HI\ intensity maps,
all to the same scale.  Top panel is the NA moment 0 map (45\as\
resolution).  The middle panel is the UN moment 0 map (25\as\
resolution).  The bottom panel shows the peak intensity (amplitude)
from the XGAUS single Gaussian fits to the UN data.  North is up and
east is to the left in all panels.  The scalebar in the bottom panel
is 2$'$ long and 20\as\ wide.

Fig.\ \ref{f:na0}. Contour plot of total \HI\ intensity from the  NA
zeroth moment map.  Contour levels are at 2.5, 7.5, 15, 25, 50 75, and
90 percent of the peak \HI\ surface brightness of 3.1 Jy beam$^{-1}$
\kms\ ($N_{\rm HI} = 1.71 \times 10^{21}\, {\rm cm^{-2}}$). In this and
the remaining maps the crosses mark the positions of fiducial stars and
the star marks the optical center from paper I, and the beam size is
shown in the lower left corner.

Fig.\ \ref{f:optun0}. 800s {\em I\/} band CCD image of NGC~2915, obtained with
the AAT F1 system overlain with \HI\ total intensity contours from the
UN moment 0 map.  Contours are at 5, 13, 25, 50, 75, and 90 percent of
the peak surface brightness of 1.39 Jy beam$^{-1}$ \kms\ ($N_{\rm HI} =
2.46 \times 10^{21}\, {\rm cm^{-2}}$).  The strong glow at lower right
in this figure and Fig.~\ref{f:optun1} is due to a bright star just
off the edge of the frame.

Fig.\ \ref{f:na1}. Velocity field from the NA first moment
map. The $V_r$ in \kms\ of the contours are indicated.

Fig.\ \ref{f:optun1}. Velocity field from the UN first moment
map overlain on the {\em I\/} band image.  The contour interval is 10
\kms.  The contour passing through the center has $V_r = 470$ \kms;
the prominent closed contours to the SE and NW of the center have $V_r
= 400$ and 540 \kms, respectively.  

Fig.\ \ref{f:globl} Global \HI\ $V_r$ profiles derived from the NA
(solid line) and UN (dotted line) data cubes.  The systemic velocity
(first moment of the profiles) are indicated with arrows of the
corresponding line type.

Fig.\ \ref{f:xgb} Line broadening, $b$, map from automatic
single Gaussian fitting using XGAUS.  The results are shown as both a
grey scale and contours at $b = 10, 20, 30, 40$ and 50 \kms.  Blank
areas are regions of either low signal to noise or complex profiles
which could not be fit with a single Gaussian. 

Fig.\ \ref{f:ring} Results of the tilted ring analysis of the three
velocity fields: NA-MOMNT, UN-MOMNT, and UN-XGAUS.  The parameters
shown are major axis position angle, $\phi$; inclination $i$, and
rotation velocity $V_{\rm rot}$. 

Fig.\ \ref{f:lv70} Contours through a position - velocity cut of the
NA data cube.  The cut is through the center of NGC2915 and at
a position angle of 70\dg, corresponding to the kinematic major axis.
The X axis indicates location along this position angle; the tick
marks are separated by 208\as.  The contours are at -3.5, 3.5, 7, 14,
21, 28, and 35 mJy/Beam.  Note the ``second rise'' in $V_{\rm rot}$
seen on both sides of the galaxy.

Fig.\ \ref{f:rresid} Grey scale plus contour plots of the residual
maps from the tilted ring fits.  The contours shown are for $\Delta
V_r = -20, -15, -10, -5, 5, 10, 15$ and 20 \kms.  Panels a, b, and c
show the NA-MOMNT, UN-MOMNT, and UN-XGAUS residuals respectively.
Note that the panels have different scales.

Fig.\ \ref{f:split} Position with respect to the center of NGC~2915
(shown as a star) of SPLOT profile measurements.  Positions marked
with small dots are well represented by single Gaussian fits, while
double Gaussians are required for the positions marked with large
dots.  The crosses mark the position of the fiducial stars shown in
Figs.~\ref{f:optun0},\ref{f:optun1}.

Fig.\ \ref{f:examprofs} Examples of SPLOT Gaussian fits (continuous
lines) to \HI\ line profiles (histogram style lines).  The profiles
were chosen to illustrate (a) a typical high signal/noise outer
profile; (b) a typical low signal/noise outer profile; (c) a broad
single profile near the center; (d) a split profile near the center;
(e) a split profile in the outer disk; (f) an atypical profile: a
narrow core on top of a normal single profile.  The positions of the
profiles in $\pm{\rm E},\pm{\rm N}$ offsets (in arcsec) from the
center of NGC~2915 are: (a) --295,101; (b) --25,311; (c) 5,11; (d)
35,--19; (e) --265,--79; (f) --325,41.

Fig.\ \ref{f:compw} Comparison of line of sight velocity dispersion,
$b$, measurements.  Azimuthal averages of the UN-MOMNT, and UN-XGAUS
measurements are shown as filled squares and triangles respectively.
Individual SPLOT measurements with signal/noise $> 3$ are shown as
$\times$'s, lower signal/noise splot measurements are indicated with
dots.  

Fig.\ \ref{f:asymd} The top panel shows the line of sight line
broadening, $b$, as a function of radius, while the middle panel shows
the gas surface density, $\Sigma_g$.  The filled circles show the data
which are azimuthal averages extracted from the UN-XGAUS $b$ map (top)
and UN zeroth moment maps (center) using concentric rings with $\phi =
-60.8$, $i=59.6$.  The broken lines show the fits to these data as
given in the text.  The contribution of the $b$ and $\Sigma$ gradients
to the total asymmetric drift correction, $\sigma_D$ are shown in the
bottom panel with the corresponding line style.  The total $\Sigma_D$
curve is shown as the solid line.

Fig.\ \ref{f:mm} The rotation curve with mass model fits.  Triangles
represent results from the UN-XGAUS data, circles from the UN-MOMNT
data, and squares from the NA-MOMNT data.  Closed symbols are $V_c$
and open symbols \Vrot.  The Holmberg radius is indicated by the
arrow.  The thick solid line in each panel shows the model fit, and
the broken lines show the contribution of the individual components:
stellar disk, dotted line; neutral gas, long dashed line; DM halo,
short dashed line.  See the text for a description of the individual
models.

Fig.\ \ref{f:optun2}  Contours of line of sight velocity dispersion
$b$ overlain on the \Halpha\ image of Marlowe \etal\ (1994).  The $b$
contours correspond to those shown in fig.~\ref{f:xgb}.  The scale bar
at lower right is 30\as\ long.

Fig.\ \ref{f:tfr} The Tully-Fisher relationship for nearby disk
galaxies compared with DDO154 and NGC 2915.  $\Delta V_{21}^i$ is the
inclination corrected \HI\ line width at 20\%\ of the peak value, and
$M_B$ is the absolute $B$ band magnitude.  The solid circles are data
from Sandage \&\ Tamman (1976); the $\times$'s are from Table 2 of
Sandage (1988).  Galaxies with $i < 40\dg$ have been excluded from
both samples and we have taken $M_B = M_{\rm pg} + 0.1$ where
necessary.

Fig.\ \ref{f:surfden} {\it Upper panel\/}: radial variation of
logarithmic surface densities.  The neutral gas measurements (solid
line) are azimuthal averages using the UN data for $R \leq 10$ kpc
(390\as) and the NA data for $R > 10$ kpc.  The projected DM density
(dashed line) is that derived from model D.  The stellar profile
(dotted line) is the {\em B\/} band surface brightness profile (paper
I) normalized to the ${\cal M}/L_B$ ratio of model D.  The critical density
required for efficient star formation (\markcite{Kennicutt, 1989}; dot
dashed line) is derived from the rotation curve and the $b$ fit of
eq.~\ref{e:bfit}. {\it Lower panel\/}: logarithm of surface densities
normalized by $\Sigma_{\rm DM}$.  Line styles are the same as for the
upper panel.

Fig.\ \ref{f:skiggyi} 800 $s$ {\em I\/} band CCD image of \skiggy, obtained
with the AAT F1 system. North is up East is to the left and the image
is 500\as\ across in each dimension.

\clearpage

\begin{planotable}{lrrr}
\tablecaption{HI Observing Log\label{t:radlog}}
\tablewidth{0pc}
\tablehead{\colhead{config.} & \colhead{UT date} & 
\colhead{baseline} & \colhead{on source} \\
\colhead{name} & & \colhead{range\tablenotemark{a}} & \colhead{(hrs)}}
\startdata
0.375  &  3 May, 1992 &  31 --  459 & 11.34 \nl
0.75 C & 22 Jan, 1993 &  46 --  750 & 10.52 \nl
1.5  D & 12 Mar, 1993 & 107 -- 1439 & 10.80 \nl
\tablenotetext{a}{In meters, excluding baselines with antenna 6.}
\end{planotable}

\begin{planotable}{l c c l}
\tablecaption{Dataset properties\label{t:HIprop}}
\tablewidth{160mm}
\tablehead{\colhead{Quantity} & \colhead{NA value} & \colhead{UN value} & \colhead{units}} 
\startdata
beam \Wfo                & $45\times 45$ & $27\times 23$ & ${\rm arcsec\times arcsec}$ \nl
pixel size               & $14\times 14\times 6.62$ & $10\times 10\times 3.31$ & 
 ${\rm arcsec\times arcsec \times \kms}$ \nl
noise/channel            & 2.9           & 3.0           & $10^{-3}\,{\rm Jy\, Beam^{-1}}$ \nl
$\int S dV$              & 145           & 139           & Jy \kms \nl
${\cal M}_{\rm HI}$             & 9.58          & 9.21          & $10^8\,\Msun$ \nl
\Wfo(global)             & 151           & 146           & \kms \nl
\Wto(global)             & 170           & 163           & \kms \nl
\Vsys(global)            & 471           & 468           & \kms \nl
\Vsys(dynamical)         & 467           & 469           & \kms \nl
$R_{\rm HI}$             & 14.9          & \nodata       & kpc \nl 
$N_{\rm HI}$(max)        & 0.88          & 1.26          & $10^{21}\,{\rm cm^{-2}}$ \nl
${\cal M}_{\rm HI}/L_B$         & 2.7           & 2.6           & $(M/L_B)_\odot$ \nl
$R_{\rm HI}/R_{\rm Ho}$   & 5.2           & \nodata      &  \nl
$R_{\rm HI}/\alpha_B^{-1}$ & 22.6         & \nodata      &  \nl
\end{planotable}

\clearpage

\begin{planotable}{rccccc}
\tablecaption{Results of tilted ring analysis. \label{t:rot}}
\tablewidth{13cm}
\tablehead{\colhead{R} & \colhead{$\phi$} & \colhead{$i$} & \colhead{\Vrot} & 
\colhead{$\sigma_D$} & \colhead{$V_c$} \\
\colhead{(\as)} & \colhead{(\dg)} & \colhead{(\dg)} & \colhead{(\kms)} & \colhead{(\kms)} & \colhead{(\kms)}}
\startdata
 27.5 & $-28 \pm 12$    & $73 \pm 24$    & $29 \pm 17   $ & 26.5 & 39 \nl 
 52.5 & $-28 \pm 11$    & $78.0 \pm 7.1$ & $37 \pm 22   $ & 38.1 & 53 \nl 
 77.5 & $-48.1 \pm 2.1$ & $63 \pm 11$    & $53 \pm 27   $ & 37.7 & 65 \nl 
102.5 & $-56.3 \pm 0.8$ & $74.8 \pm 5.4$ & $65.5 \pm 2.0$ & 31.5 & 72.7 \nl 
127.5 & $-56.8 \pm 3.9$ & $73.1 \pm 9.7$ & $71.9 \pm 9.0$ & 24.6 & 76.0 \nl 
152.5 & $-55.7 \pm 4.2$ & $70.3 \pm 8.4$ & $76.8 \pm 6.7$ & 19.0 & 79.2 \nl 
177.5 & $-53.7 \pm 0.6$ & $62.2 \pm 2.3$ & $79.9 \pm 1.3$ & 15.2 & 81.4 \nl 
202.5 & $-53.0 \pm 0.6$ & $56.0 \pm 2.7$ & $81.5 \pm 1.0$ & 12.6 & 82.5 \nl 
227.5 & $-56.3 \pm 0.5$ & $53.8 \pm 1.4$ & $82.1 \pm 4.6$ & 11.2 & 82.8 \nl 
252.5 & $-58.6 \pm 1.7$ & $55.2 \pm 1.4$ & $79.7 \pm 4.6$ & 10.5 & 80.4 \nl 
277.5 & $-61.4 \pm 1.9$ & $53.0 \pm 1.2$ & $79.7 \pm 3.7$ & 10.5 & 80.4 \nl 
302.5 & $-62.8 \pm 2.1$ & $52.2 \pm 1.1$ & $80.2 \pm 1.6$ & 10.8 & 80.9 \nl 
327.5 & $-64.0 \pm 1.0$ & $53.1 \pm 1.5$ & $79.9 \pm 0.7$ & 11.3 & 80.7 \nl 
352.5 & $-65.0 \pm 1.1$ & $53.8 \pm 3.1$ & $79.9 \pm 1.5$ & 11.9 & 80.8 \nl 
377.5 & $-65.1 \pm 1.0$ & $54.6 \pm 2.3$ & $81.0 \pm 0.7$ & 12.5 & 82.0 \nl 
412.5 & $-66.3 \pm 0.4$ & $56.2 \pm 2.0$ & $81.2 \pm 3.2$ & 13.3 & 82.3 \nl 
457.5 & $-67.5 \pm 1.6$ & $58.4 \pm 0.7$ & $82.6 \pm 2.1$ & 14.3 & 83.8 \nl 
502.5 & $-68.5 \pm 2.2$ & $58.7 \pm 1.9$ & $84.2 \pm 1.5$ & 15.2 & 85.5 \nl 
547.5 & $-68.8 \pm 0.9$ & $58.9 \pm 2.3$ & $87.0 \pm 0.5$ & 16.1 & 88.5 \nl 
592.5 & $-68.8 \pm 0.5$ & $58.2 \pm 1.1$ & $92.3 \pm 1.3$ & 16.9 & 93.9 \nl 
\end{planotable}

\clearpage

\begin{planotable}{lrrrr}
\tablecaption{Mass models\label{t:mm}}
\tablewidth{0pc}
\tablehead{\colhead{Model:} & \colhead{A} & \colhead{B} &
\colhead{C} & \colhead{D}} 
\startdata
\underline{Fit parameters} : \nl
\zsp ${\cal M}/L_B$ (solar units)     & $<0.4$                 &  $<0.14$               & $0.75 \pm 0.40$ & 1.2             \nl
\zsp $R_c$ (kpc)                      & $0.78^{+0.17}_{-0.07}$ & $1.24^{+0.19}_{-0.09}$ & $2.5 \pm 0.3$   & $1.23 \pm 0.15$ \nl
\zsp $\rho_0~{\rm (\Msun pc^{-3})}$   & $0.24 \pm 0.06$        & $0.10 \pm 0.02$        & $0.07 \pm 0.02$ & $0.10 \pm 0.02$ \nl
\zsp $\sigma_0$ (\kms)                & $39.9 \pm 0.8$         & $41.2 \pm 0.9$         & $51.9 \pm 0.9$  & $41.0 \pm 1.1$  \nl
\zsp $V_\infty$ (\kms)                & $88 \pm 2$             & $91 \pm 2$             & $73 \pm 1$      & $91 \pm 2$      \nl
\zsp rms (\kms)                       &     2.8                &  3.8                   &     4.1         &        3.8      \nl
\zsp Reduced $\chi^2$                 &    1.01                & 1.87                   &    2.14         &       1.80      \\[2mm]
\underline{At \RHo\ (2.9 kpc)} : \nl 
\zsp $V_c$ (\kms)                                  & 73   &  66 &  73  & 71  \nl
\zsp ${\cal M}_T ~(10^9\, \Msun)$                  & 3.6  & 3.0 & 3.6  & 3.4 \nl
\zsp ${\cal M}_T/L_B$ (solar units)                & 10.1 & 8.4 & 10.2 & 9.6 \nl
\zsp ${\cal M}_{\rm dark}/{\cal M}_{\rm Luminous}$ & 27   &  22 &  8.3 & 5.2 \\[2mm]
\underline{Within last point (15.2 kpc)} : \nl
\zsp $V_c$ (\kms)                                  & 87   &  87 &  81  & 88   \nl
\zsp ${\cal M}_T ~(10^9\, \Msun)$                  & 27   &  27 &  23  & 27   \nl
\zsp ${\cal M}_T/L_B$ (solar units)                & 74   &  75 &  65  & 76   \nl
\zsp ${\cal M}_{\rm dark}/{\cal M}_{\rm Luminous}$ & 28   &  28 &  19  & 19   \\[2mm]
\tablecomments{Errors are 90\%\ confidence limits and upper
limits are 95\%\ confidence limits.  They were determined from
Monte-Carlo simulations assuming uniform $V_c$ and $V_{\rm rot}$ errors
of $\pm 3$ \kms.}
\end{planotable}

\clearpage

\begin{figure}[h]
\caption{}\label{f:cube}
\end{figure}

\begin{figure}[h]
\caption{}\label{f:morph}
\end{figure}

\begin{figure}[h]
\caption{}\label{f:na0}
\end{figure}

\begin{figure}[h]
\caption{}\label{f:optun0}
\end{figure}

\begin{figure}[h]
\caption{}\label{f:na1}
\end{figure}

\begin{figure}[h]
\caption{}\label{f:optun1}
\end{figure}

\begin{figure}[h]
\caption{}\label{f:globl}
\end{figure}

\begin{figure}[h]
\caption{}\label{f:xgb}
\end{figure}

\begin{figure}[h]
\caption{}\label{f:ring}
\end{figure}

\begin{figure}[h]
\caption{}\label{f:lv70}
\end{figure}

\begin{figure}[h]
\caption{}\label{f:rresid}
\end{figure}

\begin{figure}[h]
\caption{}\label{f:split}
\end{figure}

\begin{figure}[h]
\caption{}\label{f:examprofs}
\end{figure}

\begin{figure}[h]
\caption{}\label{f:compw}
\end{figure}

\begin{figure}[h]
\caption{}\label{f:asymd}
\end{figure}

\begin{figure}[h]
\caption{}\label{f:mm}
\end{figure}

\begin{figure}[h]
\caption{}\label{f:optun2}
\end{figure}

\begin{figure}[h]
\caption{}\label{f:tfr}
\end{figure}

\clearpage

\begin{figure}[h]
\caption{}\label{f:surfden}
\end{figure}

\begin{figure}[h]
\caption{}\label{f:skiggyi}
\end{figure}

\end{document}